\documentclass[twocolumn,prl,superscriptaddress]{revtex4}
\usepackage{amsfonts}
\usepackage{amssymb}
\usepackage{amsmath}
\usepackage{epsfig}
\usepackage{color}
\usepackage{graphics, graphicx}
\usepackage{bbold}
\usepackage{psfrag}
\usepackage{mathcomp}
\usepackage{subfigure}
\usepackage{verbatim}
\usepackage[colorlinks,citecolor=blue]{hyperref}

\makeatletter

\newcommand{\Rmnum}[1]{\expandafter\@slowromancap\romannumeral #1@}
\makeatother

\begin{document}
	
\title{Dynamical Detection of Topological Spectral Density}
	
	\author{Jia-Hui Zhang}
	\affiliation{State Key Laboratory of Quantum Optics and Quantum Optics Devices, Institute
		of Laser Spectroscopy, Shanxi University, Taiyuan, Shanxi 030006, China}
	\affiliation{Collaborative Innovation Center of Extreme Optics, Shanxi
		University, Taiyuan, Shanxi 030006, China}

	\author{Feng Mei}
	\email{meifeng@sxu.edu.cn}
	\affiliation{State Key Laboratory of Quantum Optics and Quantum Optics Devices, Institute
		of Laser Spectroscopy, Shanxi University, Taiyuan, Shanxi 030006, China}
	\affiliation{Collaborative Innovation Center of Extreme Optics, Shanxi
		University, Taiyuan, Shanxi 030006, China}
	
	\author{Liantuan Xiao}
	\affiliation{State Key Laboratory of Quantum Optics and Quantum Optics Devices, Institute
		of Laser Spectroscopy, Shanxi University, Taiyuan, Shanxi 030006, China}
	\affiliation{Collaborative Innovation Center of Extreme Optics, Shanxi
		University, Taiyuan, Shanxi 030006, China}
	
	\author{Suotang Jia}
	\affiliation{State Key Laboratory of Quantum Optics and Quantum Optics Devices, Institute
		of Laser Spectroscopy, Shanxi University, Taiyuan, Shanxi 030006, China}
	\affiliation{Collaborative Innovation Center of Extreme Optics, Shanxi
		University, Taiyuan, Shanxi 030006, China}

	\date{\today}
\begin{abstract}
Local density of states (LDOS) is emerging as powerful means of exploring classical-wave topological phases. However, the current LDOS detection method remains rare and merely works for static situations. Here, we introduce a generic dynamical method to detect both the static and Floquet LDOS, based on an elegant connection between dynamics of chiral density and local spectral densities. Moreover, we find that the Floquet LDOS allows to measure out Floquet quasienergy spectra and identify topological $\pi$ modes. As an example, we demonstrate that both the static and Floquet higher-order topological phase can be universally identified via LDOS detection, regardless of whether the topological corner modes are in energy gaps, bands or continue energy spectra without bandgaps. Our study opens a new avenue utilizing dynamics to detect topological spectral densities and provides a universal approach of identifying static and Floquet topological phases.
\end{abstract}
	\maketitle

\emph{Introduction}. Topological phases (TPs) are hallmarked by the appearance of topological boundary modes (TBMs). Consequently, spectral measurements of TBMs have become a mainstream approach of identifying classical-wave TPs~\cite{TPrev2014,TPrev2016,TPrev2017,TPrev2018,TPrev2019a,TPrev2019b,TPrev2022a,TPrev2022b,
Deng2023,Zhu2023}. Despite its success, this approach only works for the TPs with TBMs residing in energy gaps, and can not apply to the TPs with TBMs embedded into energy bands~\cite{2020scienceBahl}. The latest development is that, by detecting local spectral density, i.e., local density of states (LDOS)~\cite{2020scienceBahl}, one can identify the TPs with in-band TBMs~\cite{2020scienceBahl,2022Xie}. Beyond this, the LDOS detection has also promoted the observation of bulk-defect correspondences~\cite{2021natureJiang,2022NcBahl,2020PRappliBahl,2021natureBahl} and detection of fractional charges~\cite{2020scienceBahl,2023PRappliLu,Jiangarxiv,2023APLLu,2023SciBullLIu}. However, the current classical-wave LDOS detection means remain rare, as reported recently in photonic systems ~\cite{2021natureBahl,2021natureJiang,2020scienceBahl} by measuring reflection spectroscopy and in acoustic systems by measuring volume flow rate~\cite{2023PRappliLu}. Moreover, both methods can not apply to detect the LDOS of time-dependent periodically driven TPs. Therefore, it is highly valuable to develop new LDOS detection approaches.

Meanwhile, Floquet TPs arising from periodic driving recently have attracted significant interests in classical-wave systems~\cite{2013Rechtsman,2014Chong,2016Chonga,2016Zhang,2016Alu,2016Zhu,2016Segev,2016Chong,2016Carusotto,
	2017Szameit,2017Thomson,2019Zhen,2019Ni,2019Ivanov, 2020Van,2020Yang,2020Szameit,2020Rechtsman,2020Maczewsky,2020Segev,2020Ivanov,2021Rechtsman,2022Van,2022Zhen,Pan2020a,Pan2020b,
	2022Chong,2022Szameit,2022Biesenthal,2022Pan,2022Jianwei,2019Cheng,2021Lia,2021Lib,2022Zhang,2022Gong}. Distinct from static TPs, Floquet TPs are classified through quasienergy spectra~\cite{FTP2010a,FTP2010b}, which are $2\pi$ periodic, leading to a plethora of unique topological properties~\cite{FTP2010a,FTP2010b,Jiang2011,FTP2011,Pi2022,2013Runder,2013An,2014Delplace,2014Gong,
2016Fruchart,2019Seradjeh,2020Hua,2020Hub,2020Peng,2021Gong,2021An,Fpi2022a}, such as the topological $\pi$ modes pinned to half of the driving frequency. Nevertheless, how to measure the periodic quasienergy spectra of Floquet topological systems~\cite{2013Rechtsman,2014Chong,2016Chonga,2016Zhang,2016Alu,2016Zhu,2016Segev,2016Chong,2016Carusotto,
	2017Szameit,2017Thomson,2019Zhen,2019Ni,2019Ivanov, 2020Van,2020Yang,2020Szameit,2020Rechtsman,2020Maczewsky,2020Segev,2020Ivanov,2021Rechtsman,2022Van,2022Zhen,Pan2020a,Pan2020b,
	2022Chong,2022Szameit,2022Biesenthal,2022Pan,2022Jianwei,2019Cheng,2021Lia,2021Lib,2022Zhang,2022Gong} and spectrally detect the topological $\pi$ modes, even if they are within the energy gaps, remains a challenging question~\cite{SYZ2023}.

\begin{figure}[t]
		\includegraphics[scale=0.45]{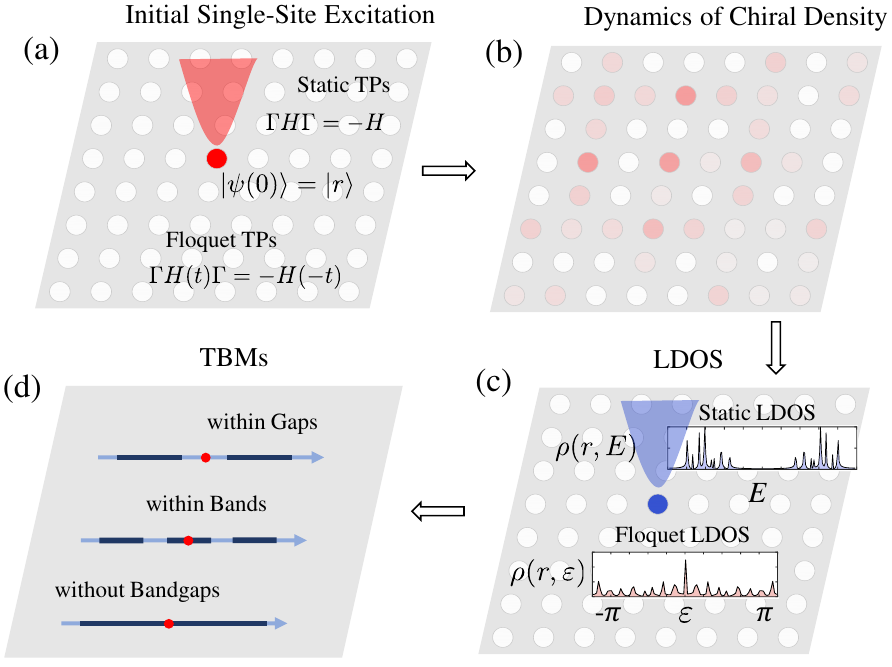}
		\caption{Dynamics of CD is connected with static and Floquet LDOS, which respectively allows for universally identifying the static and Floquet TPs, regardless of whether the static and Floquet TBMs are in the energy gaps, bands or the continue energy spectra without bandgaps. }
		\label{Fig1}
\end{figure}

In this Letter, we introduce a dynamical method to address all these challenging questions. We uncover that there exists an elegant fundamental connection between the dynamics of chiral density (CD) and the spectral density in both static and periodically driven systems. This connection leads to an universal approach of detecting both the static and Floquet LDOS (Fig. \ref{Fig1}), simply from the time evolutions of photonic intensities or acoustic pressures. We also find that both the periodic quasienergy spectra and the topological $\pi$ modes can be directly detected by the Floquet LDOS. As exemplified in probing static and Floquet higher-order TPs (HOTPs), our LDOS approach can universally identify the static and Floquet topological corner modes (TCMs), regardless of whether they are in the energy gaps, bands or the continue energy spectra without bandgaps (Fig. \ref{Fig1}), remarkably even can identify the TCMs embedded in a continue quasienergy spectrum which resembles the one in the trivial case (Figs. \ref{Fig3}(g,j)). Beyond previous methods detecting the spatial distributions of TBMs, our LDOS approach allows to gain deep insights, including their spectral signatures, numbers and fractional charges. Particularly for the Floquet HOTPs, our approach is capable to detect the Floquet quasienergy spectra, the $\pi$-energy topological corner modes ($\pi$-TCMs) and their discriminations from the 0-TCMs, the numbers of 0- and $\pi$-TCMs, and even the recently discovered fractional topological $\pi$ modes, all of which can not be achieved by previous methods. Our approach also has a wide range of applications and can be used in general classical-wave and quantum chiral systems for either static and Floquet cases.

\emph{Relationship between dynamics of CD, Loschmidt amplitude and LDOS}. We start by studying the dynamics of CD in a chiral TP that belongs to an important class of category in the classification table of TPs. The chiral symmetry guarantees $\Gamma^{-1} H\Gamma=-H$, naturally which could reverse forward into backward time evolution,
\begin{equation}
\Gamma^{-1} U(t)\Gamma=U(-t).
\end{equation}
Implementing time-reversed evolution is vital for studying quantum information scrambling~\cite{2019Rey} and Loschmidt echo~\cite{2020Echo}. However, the consequence of chiral-symmetry-enabled time-reversed operation has not been revealed before. Below, we report a finding benefiting from such evolution. Consider the initial system state as a single-site state $|\psi(0)\rangle=|r\rangle$, with photons initially inputting into the lattice site $r$. This state is also the eigenstate of chiral operator, satisfying $\Gamma |\psi(0)\rangle=\lambda_r|\psi(0)\rangle$, with $\lambda_r=\pm1$ being the eigenvalues. After its time evolution and making use of the time-reversed operation, we find a correspondence between the CD at $t/2$ and the Loschmidt amplitude at $t$ (see the Appendix), i.e.,
\begin{align}
\left<\Gamma(t/2)\right>_r=\lambda_r\langle\psi(0)|\psi(t)\rangle=\lambda_rG(t).
\label{cd}
\end{align}
Via this correspondence, experimental detection of Loschmidt amplitudes $G(t)$ becomes much simpler, without requiring the full information of lattice wave functions. This result is of great value for utilizing classical-wave systems to explore dynamical topological phase transitions~\cite{2018Heyl} and can motivate broad future interests.

\begin{figure*}
		\includegraphics[scale=0.53]{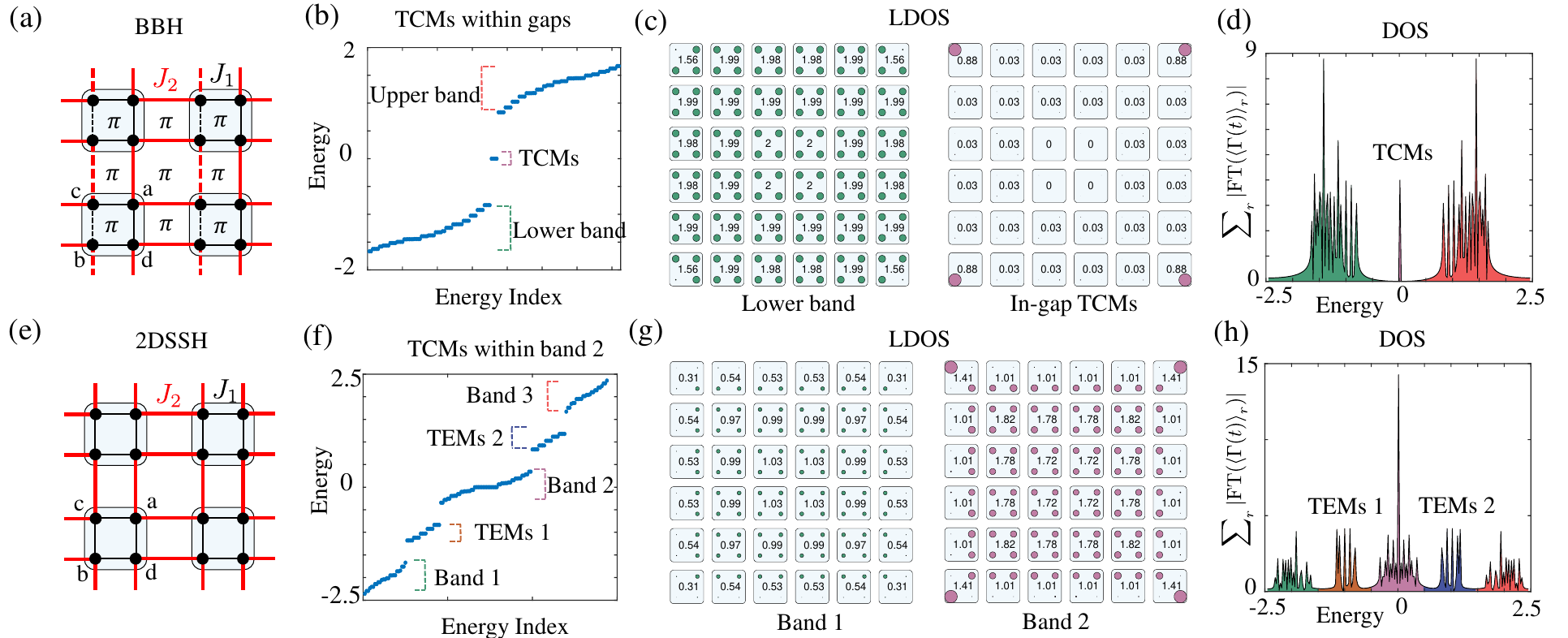}
		\caption{Dynamical detection of static LDOS for identifying HOTPs in the (a) BBH and (e) 2DSSH models. (b) OBC energy spectra of the BBH model in nontrivial HOTPs. (c) Spatial distributions of static LDOS integrated over the lower band and 0-energy gap. (d) Sum of LDOS over all sites. Each site is represented as a circle with radius proportional to the LDOS. Each number in the box gives the LDOS values in the corresponding unit cell. (f-h) Same as (b-d) but for nontrivial HOTPs in the 2DSSH model. The parameter is $J_2=5J_1$ and the evolution time is $t=20/J_{2}$.}
		\label{Fig2}
\end{figure*}

\begin{figure*}
		\includegraphics[scale=0.53]{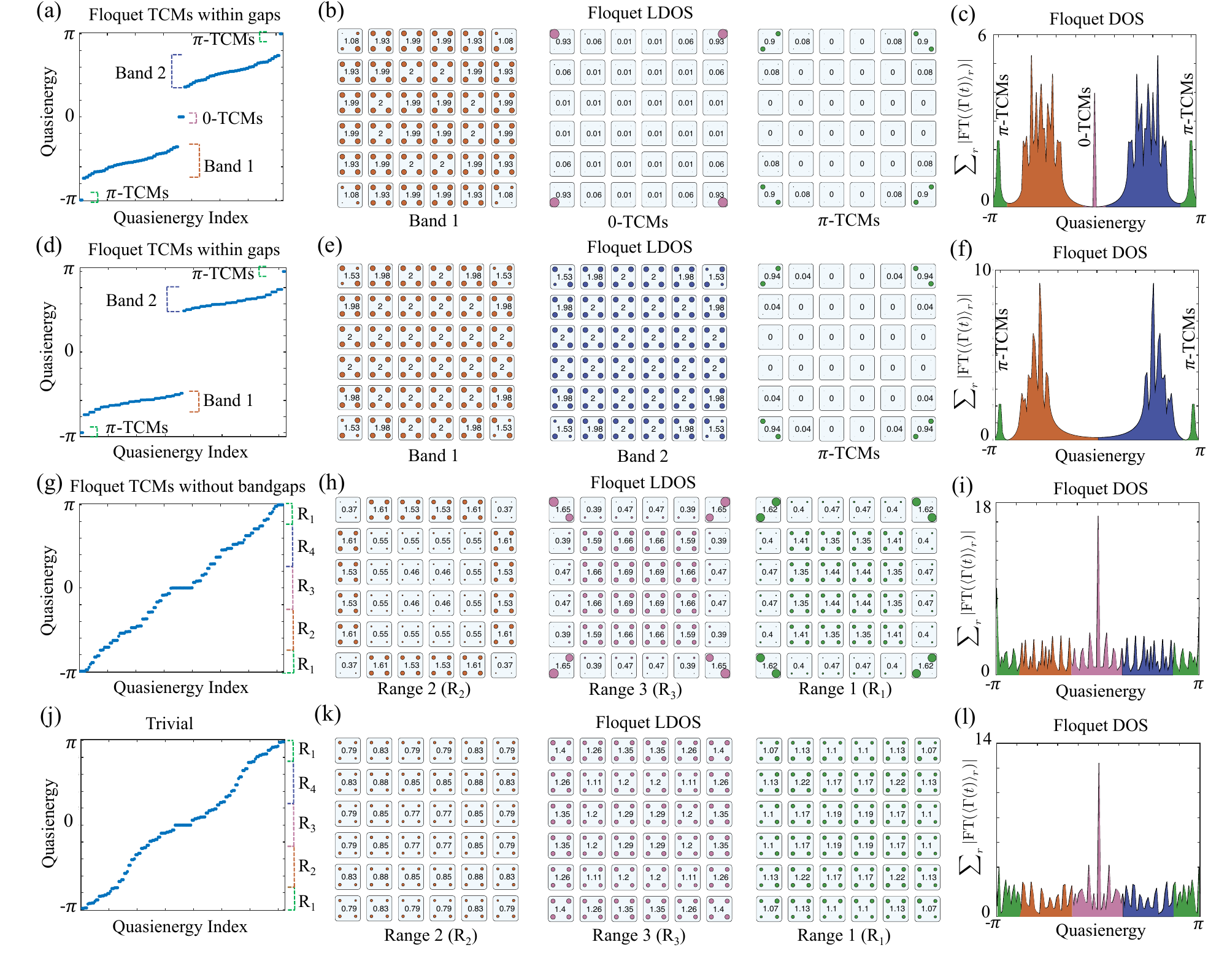}
		\caption{Dynamical detection of Floquet LDOS for identifying Floquet HOTPs. (a,d) OBC quasienergy spectra of the periodically driven BBH model respectively in two different nontrivial Floquet HOTPs. (b, e) show the respective Floquet LDOS integrated over different spectral ranges. (d, f) sum of LDOS over all sites. (g-l) Same as (a-f) but for the periodically driven 2DSSH model. (i,l) correspondingly present the spatial distributions of Floquet LDOS integrated over the spectra ranges $R_{1,2,3}$. The parameters are (a-c) $J_1T=0.63\pi$ and $J_2T=1.55\pi$, (d-f) $J_1T=1.27\pi$ and $J_2T=0.56\pi$, (g-i) $J_1T=0.8\pi$ and $J_2T=1.7\pi$, (j-l) $J_1T=0.8\pi$ and $J_2T=0.46\pi$, and $N=50$. }
		\label{Fig3}
	\end{figure*}

Our target is to establish a connection between the dynamics of CD and the spectral density. Suppose the eigenvalues and eigenstates of $H$ are $E_m$ and $|\psi_m\rangle$, respectively. Expanding the Loschmidt amplitude $G(t)$ in the eigenbasis $\{|\psi_m\rangle\}$, Eq. (\ref{cd}) becomes $\left<\Gamma(t/2)\right>_r=\lambda_r\textstyle\sum_m |c^m_{r}|^2e^{-iE_m t}$, where $c^m_{r}=\langle\psi_m|r\rangle$. After performing a discrete Fourier transformation (FT), we obtain an elegant connection between the dynamics of CD (or the Loschmidt amplitude) and the LDOS in the lattice site $r$ at the energy $E$, i.e.,
\begin{align}
	\rho(r,E)=\text{FT}[\left<\Gamma(t/2)_r\right>]=\lambda_r\textstyle\sum_m |c^m_{r}|^2 \delta(E-E_m),
\label{ldos}
\end{align}
which has not been found before, opening a much neater way to directly detect the local spectral density. Notice that, if the initial state is not the chiral eigenstate, $\left<\Gamma(t/2)\right>_r$ is related to $e^{-i(E_m-E_n)t}$, then we can not exactly extract the spectral density at $E_m$. As exemplified in Fig. \ref{Fig2}, dynamically detecting LDOS allows us to unambiguously distinguish the nontrivial TPs, even the TBMs are in the bands.

\emph{Generalized to Floquet LDOS in periodically driven systems}. Besides static TPs, dynamics of CD can also detect both the LDOS and the quasienergy spectra of periodically driven topological systems $H(t)=H(t+T)$. The Floquet topology is characterized by Floquet operator $U_F(T)=\mathcal{T}e^{-i\int_{0}^{T}H(t)dt}$, with its energy spectrum given by
	\begin{equation}
		U_F(T)|\phi_m\rangle=e^{-i\varepsilon_mT}|\phi_m\rangle,
	\end{equation}
where the eigenvalues $\{\varepsilon_m\}$ form the quasienergy spectrum and the eigenstates $\{|\phi_m\rangle\}$ constitute the Floquet eigenstates. Intriguingly, the quasienergy spectrum $\varepsilon$ is periodic, ranging from $-\pi$ to $\pi$, with period $2\pi$ in units of $1/T$. Furthermore, as the Floquet system satisfies chiral symmetry, i.e., $\Gamma H(t)\Gamma^{-1}=-H(-t)$, which leads to $\Gamma U_F(nT)\Gamma^{-1}=U_F^{-1}(nT)$, with $n$ being the number of driving periods. As a result, a connection between the Floquet LDOS and the CD at $t=nT/2$ is established, i.e.,
\begin{align}
	\rho(r,\varepsilon)=\text{FT}[\left<\Gamma(nT/2)_r\right>]=\lambda_r\textstyle\sum_m |c^m_{r}|^2 \delta(\varepsilon-\varepsilon_m),
\label{fldos}
\end{align}
where $c^m_{r}=\langle\phi_m|r\rangle$. As exhibited in Fig. \ref{Fig3}, Floquet LDOS can directly map out the periodic quasienergy spectra and identify the unique topological $\pi$ modes, even the Floquet TBMs are within a continue quasienergy spectra without bands and gaps.

\emph{Application to static HOTPs with TCMs within gaps and bands}. Our method can identify the recently highly concerned
HOTPs~\cite{BBH,2017Brouwer,2017Fang,2018Neupert,2018Lee,2021Xie,2020Kim,2018Bahl,2018serra,2018Rechtsman,
2019Zhang,2019Khanikaev,2019Hassan,2019Jiang,2019Chen,2019Lu,2019Dong,2019Hafezi,2019Xue,2020Qiu,2020Ni,2020Xie,2021Liu,
2020Jiangjh,2021Wangyao,2021Qiu,2023Chen,2022Zheng,2022Yangzj,2022Bahl,2023Gao}, regardless of whether the TBMs are in energy gaps or bands. As illustrated in Figs. \ref{Fig2}(a,e), we exemplify our method based on the paradigmatic Benalcazar-Bernevig-Hughes (BBH)~\cite{BBH} and two-dimensional Su-Schrieffer-Heeger (2DSSH)~\cite{SSH} models. The two models and their periodically driven counterparts (next section) feature same chiral symmetry operators, defined as $\Gamma=\textstyle\sum_{x,y}a_{x,y}^\dagger a_{x,y}+b_{x,y}^\dagger b_{x,y}-c_{x,y}^\dagger c_{x,y}-d_{x,y}^\dagger d_{x,y}$. The specific procedure for chiral density detection is presented as follows. Firstly, using lasers to excite the single-waveguide at the lattice site $r$, preparing the system into the single-site state $|\psi(0)\rangle=|r\rangle$, which is exactly the eigenstate of the chiral symmetry operator. Next, controlling the propagation distance of photons in the waveguide lattice to settle the evolution time, and measuring the corresponding photon intensity distributions. With the measured photon density in each waveguide at different time, one can obtain the chiral density $\langle\Gamma(t)\rangle_r=\textstyle\sum_{x,y}\langle n_{a_{x,y}}(t)\rangle_r+\langle n_{b_{x,y}}(t)\rangle_r-\langle n_{c_{x,y}}(t)\rangle_r-\langle n_{d_{x,y}}(t)\rangle_r$, where $n_s$ denotes the photon number operator at the waveguide sublattice $s$ in the unit cell $(x,y)$. Notably, this detection procedure is quite common in practical waveguide lattice platforms~\cite{2013Rechtsman}.

The open-boundary-condition (OBC) energy spectrum of the BBH model (Fig. \ref{Fig2}(b)) shows that, besides the two degenerate bands, there emerges four in-gap 0-energy topological corner modes (0-TCMs). Fig. \ref{Fig2}(c) presents the spatial maps of the LDOS extracted from the dynamics of CD integrated over different energy range.  As shown, the lower-band LDOS in the four corner unit cells are no longer 2 but closing to 1.5, as a result of the in-gap corner modes occupying there, with the LDOS difference manifesting the corner modes. The four degenerate 0-TCMs can also be directly observed through the in-gap LDOS. As shown, it is maximally localized in the four corner unit cells. Summing the in-gap LDOS over all sites gives the number of 0-TCM $N_0=4.36$, which can be very closing to 4 for longer evolution time and larger lattice size. Fig. \ref{Fig2}(d) exhibits that the OBC energy spectrum is precisely measured out by the DOS, i.e. the sum of LDOS over all sites. The distinct peak at zero energy clearly detect the in-gap 0-TCMs.

In contrast, there is no 0-energy gap in the nontrivial 2D SSH model (Fig. \ref{Fig2}(f)), where the 0-TCMs are embedded into the band 2 and the topological edge modes (TEMs) appear at the edges of band 2. Consequently, we can not recognize the 0-TCMs from the energy spectrum. However, the LDOS can achieve this purpose (Fig. \ref{Fig2}(g)). As shown, the band-2 LDOS is dominated in four corner sites, hallmarking the existence of four 0-TCMs in this band. The LDOS integrated over the TEMs also identifies the edge states. The fraction corner charge carried by the 0-TCM is extracted from the LDOS of band 1 via $Q_c=\rho-(\sigma_1+\sigma_2)\,\,\text{mod}\,\,1$~\cite{2020scienceBahl}, where $\rho$ and $\sigma_{1,2}$ are respectively the LDOS at the corner and two edges. Taking an average over four corners, we obtain $Q_c=0.24$ that is very closing to the theoretical prediction $\frac{1}{4}$~\cite{FCC}. Fig. \ref{Fig2}(h) confirms that the energy spectra is measured by the DOS (the sum of LDOS over all sites).

\emph{Application to Floquet HOTPs with $0$- and $\pi$-TCMs within gaps and without bandgaps}. Now we show that our method can apply to probe periodically driven TPs, particularly to detect the Floquet HOTPs that has recently been experimentally implemented in acoustic waveguide lattices~\cite{2022Zhang}. Without loss of generality, we consider the periodically driven BBH and 2DSSH models, with the intra- and inter-cell couplings cyclically modulated as $J_1(t)=J_1$ for $t\in\left[0,T/4\right]$ and $\left(3T/4,T\right]$, $J_2(t)=J_2$ for $t\in\left(T/4,3T/4\right]$, otherwise zero. Specifically, Figs. \ref{Fig3}(a,d) show the OBC quasienergy spectrum of the Floquet BBH model respectively for $(Z_0=1,Z_{\pi}=1)$ and $(Z_0=0,Z_{\pi}=1)$~\cite{SM}, with $Z_{0,\pi}$ determining the numbers of 0- and $\pi$-TCMs at each corner. The Floquet LDOS in Figs. \ref{Fig3}(b,e) allow us to gain deep insights into Floquet HOTPs. Fig. \ref{Fig3}(b) shows that the Floquet LDOS in the 0- and $\pi$-energy gaps are maximally located at different corner sublattices, revealing that there are one 0-TCM and one $\pi$-TCM at each corner and they can be distinguished by the different corner-sublattice-localized features. Moreover, via summing the in-gap LDOS over all sites, we obtain the numbers of 0- and $\pi$-TCMs $N_0=4.44$ and $N_{\pi}=4.24$ in Fig. \ref{Fig3}(b), and $N_{\pi}=4.08$ in Fig. \ref{Fig3}(e), both closing to 4. While for the band 1, compared to the bulk LDOS around $\rho=2$, the corner LDOS are closing to $\rho=1$, and the difference $\delta\rho=1$ manifests the 0- and $\pi$-TCMs, as the corner sublattices with decreased LDOS are those occupied by the two TCMs. This feature is further evidenced in Fig. \ref{Fig3}(e) where the reduced LDOS at each corner due to the $\pi$-TCMs are around $\delta\rho=0.5$. The Floquet DOS are respectively presented in Fig. \ref{Fig3}(c,f), as shown which measures out the entire quasienergy spectra. More importantly, the Floquet DOS can detect the smoking-gun evidence of Floquet TPs, i.e, which not only can identify the emergence of 0- and $\pi$-TCMs in the quasienergy gaps, but also can pinpoint their locations at zero and half of the driving frequencies.

Figs. \ref{Fig3}(g,j) show that the 0- and $\pi$-TCMs in the nontrivial 2DSSH model are embedded into a continue quasienergy spectrum without bandgaps. As the trivial and nontrivial quasienergy spectra have a similar look, they are undistinguishable by the Floquet DOS in Figs. \ref{Fig3}(i,l), let alone identifying the 0- and $\pi$-TCMs. As shown in Figs. \ref{Fig3}(h,k), the Floquet LDOS~\cite{R} can solve this problem. For the nontrivial case, the Floquet LDOS integrated over $R_2$ dominated in the edge unit cells reveals the existence of TEMs in this spectral range, while the Floquet LDOS integrated over $R_{1,3}$ dominated in the four corner unit cells clearly identify the 0- and $\pi$-TCMs, respectively located at quite different corner sublattices. In sharp contrast, the Floquet LDOS in the trivial case are homogeneously distributed over all unit cells.

\begin{figure}
	\includegraphics[scale=0.46]{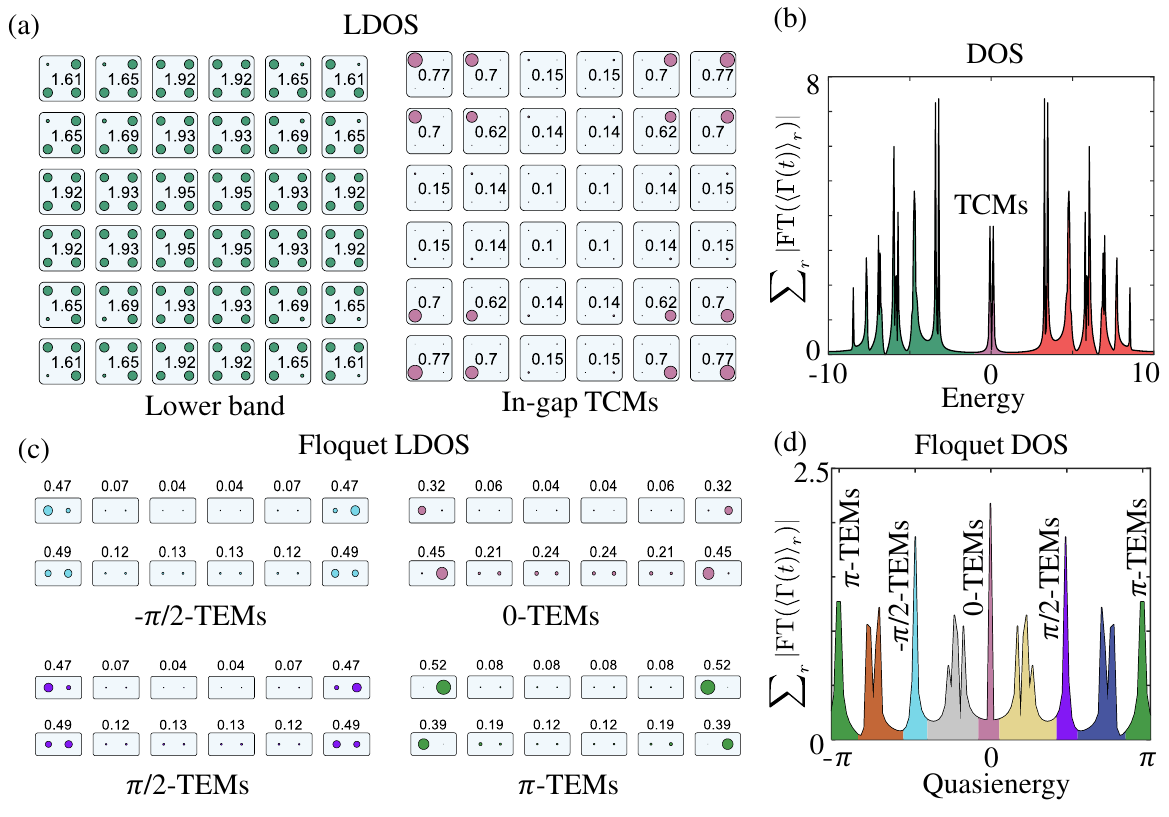}
	\caption{Dynamical detection of LDOS for identifying Z-class HOTPs and square-root Floquet TPs. (a) Spatial distributions of static LDOS integrated over the lower band and the $0$-energy gap of a Z-class HOTP with four 0-TCMs at each corner. (b) Sum of LDOS over all sites. (c) Spatial distributions of Floquet LDOS integrated over different energy gaps for a square-root Floquet TP. (d) Sum of Floquet LDOS over all sites.}
	\label{Fig4}
\end{figure}

\emph{Application to Z-class HOTPs and square-root Floquet TPs}. Our method can also probe the recently discovered cutting-edge TPs, such as the $Z$-class HOTPs~\cite{LRHOTI2022} and square-root Floquet TPs~\cite{2022Gong}. As an example, Figs. \ref{Fig4}(a,c) respectively show the LDOS of a nontrivial $Z$-class HOTP with four 0-TCMs at each corner and a nontrivial square-root Floquet TP with $\pm\pi/2$ TEMs (see~\cite{SM} for more details). As shown, the in-gap (lower-band) LDOS in Fig. \ref{Fig4}(a) is dominated (tiny) at four sites of each corner, identifying that there are four 0-TCMs at each corner. This can not be achieved from the OBC energy spectrum measured by the DOS shown in Fig. \ref{Fig4}(b). The Floquet LDOS in Fig. \ref{Fig4}(d) illustrates that, the sum of Floquet LDOS can precisely detect the smoking-gun evidence of $\pm\pi/2$ TEMs, i.e., emerged in the quasienergy gaps respectively at positive and negative quarter of the driving frequencies, which can advance further experiments on square-root Floquet TPs~\cite{2022Gong}. Moreover, the energy-resolved Floquet LDOS (Fig. \ref{Fig4}(c)) also uncovers that the spatial distributions of $\pm\pi/2$ TEMs are same and the $0$ and $\pi$ TEMs populate different sublattices.

\begin{acknowledgments}
\emph{Acknowledgment}. The authors thank for the positive comments and helpful suggestions from Biye Xie, Yiming Pan and Zhaoju Yang. This work was supported by the National Key Research and Development Program of China (Grant No. 2022YFA1404201), National Natural Science Foundation of China (NSFC) (Grant No. 12034012, 12074234), Changjiang Scholars and Innovative Research Team in University of Ministry of Education of China (PCSIRT)(IRT\_17R70), Fund for Shanxi 1331 Project Key Subjects Construction, 111 Project (D18001).
\end{acknowledgments}

\appendix
\setcounter{equation}{0}
\renewcommand\theequation{A.\arabic{equation}}

\emph{Appendix---The relationship between dynamics of CD and Loschmidt amplitude}. For a static chiral system initially excited into the single-site state $|\psi(0)\rangle =|r\rangle$, the dynamics of CD responding to such local excitation is given by
\begin{align}
	\langle\Gamma(t/2)\rangle_r&=\langle\psi(0)| U^{-1}(t/2)\Gamma U(t/2) |\psi(0)\rangle    \nonumber\\
	&=\lambda_r\langle\psi(0)|\Gamma^{-1}U^{-1}(t/2)\Gamma  e^{-iHt/2}|\psi(0)\rangle.
\end{align}
The time-reversed evolution enabled by the chiral operation is manifested in the second row, which leads to the connection with the Loschmidt amplitude
\begin{equation}
	\langle\Gamma(t/2)\rangle_r=\lambda_r\langle\psi(0)|U(t)|\psi(0)\rangle=\lambda_rG(t).
\end{equation}
Interestingly, this relationship holds even in a periodically driven system. Specifically, as the number of evolution period $n$ is even, the dynamics of CD is formulated as
\begin{align}
	\langle\Gamma(nT/2)\rangle_r&=\langle\psi(0)|U^{-1}_F(\frac{nT}{2})\Gamma U_F(\frac{nT}{2})|\psi(0)\rangle    \nonumber\\
	&=\lambda_r\langle\psi(0)|\Gamma^{-1}U^{-1}_F(\frac{nT}{2})\Gamma U_F(\frac{nT}{2})|\psi(0)\rangle,    \nonumber\\
    &=\lambda_r\langle\psi(0)|U_F(nT)|\psi(0)\rangle,   \nonumber\\
    &=\lambda_rG(nT).
\end{align}
The derivation is different for odd $n$. Notice that the Floquet operator can be rewritten as
\begin{align}
	U_F(T)=U_2U_1,
\end{align}
where $U_{1,2}$ represent the evolution operators in the two half periods. Intriguingly, we find that $U_{1,2}$ also satisfy a time-reversed relationship
\begin{align}
\Gamma^{-1} U_1\Gamma=U^{-1}_2,\,\,\,\Gamma^{-1} U_2\Gamma=U^{-1}_1
\end{align}
With these features, the CD at half-odd periods connected with the Loschmidt amplitude is proved,
\begin{align}
	&\langle\Gamma(\frac{nT}{2})\rangle_r=\langle\psi(0)|U^{-1}_F(\frac{(n-1)T}{2})U^{-1}_1\Gamma U_1U_F(\frac{(n-1)T}{2})|\psi(0)\rangle    \nonumber\\
	&=\lambda_r\langle\psi(0)|\Gamma^{-1}U^{-1}_F(\frac{(n-1)T}{2})U^{-1}_1\Gamma U_1U_F(\frac{(n-1)T}{2})|\psi(0)\rangle,    \nonumber\\
    &=\lambda_r\langle\psi(0)|U_F(nT)|\psi(0)\rangle,   \nonumber\\
    &=\lambda_rG(nT).
\end{align}


\begin{thebibliography}{81}
\bibitem{R} The integrated spectral ranges for the LDOS are divided into R$_1$: $\varepsilon \in (3\pi/4,\pi]$ and $(-\pi,-3\pi/4]$, R$ _2$: $\varepsilon \in (-3\pi/4,-\pi/4]$, R$_3$: $\varepsilon \in (-\pi/4,\pi/4]$ and R$_4$: $\varepsilon \in (\pi/4,3\pi/4]$.
\end{thebibliography}
\end{document}